\documentclass[12pt]{article}
\pdfoutput=1

\usepackage{geometry,amsmath,amsfonts}
\usepackage{slashed}
\usepackage{graphicx}
\usepackage{epstopdf}
\usepackage{mathrsfs}
\usepackage{amssymb}
\usepackage{verbatim}
\usepackage{color}
\usepackage{multirow}
\usepackage{subfig}
\usepackage{cite}
\usepackage{setspace}
\usepackage{titlesec}
\usepackage{hyperref}

\hypersetup{colorlinks=true,linkcolor=blue,citecolor=blue,urlcolor=blue}

\titleformat{\section}{\Large\bfseries}{\thesection.}{0.5em}{}

\begin{document}
\vspace*{8mm}

\begin{center}

{\large\bf Iteration on the Higgs-portal for vector Dark Matter}
 
\vspace*{3mm}

{\large\bf and its effective field theory description 
}

\vspace*{9mm}

{\sc Giorgio Arcadi}$^{1,2}$,  {\sc Juan Carlos Criado}$^3$ and {\sc Abdelhak~Djouadi}$^3$ 

\vspace*{9mm}

{\small

$^1$ Dipartimento di Scienze Matematiche e Informatiche, Scienze Fisiche e Scienze della Terra, \\ Universita degli Studi di Messina, Via Ferdinando Stagno d'Alcontres 31, I-98166 Messina, Italy.\smallskip

$^2$ INFN Sezione di Catania, Via Santa Sofia 64, I-95123 Catania, Italy. \smallskip

$^3$ CAFPE and Departamento de Fisica Te\'orica y del Cosmos,\\ Universidad de Granada, E--18071 Granada, Spain.\smallskip

}

\end{center}

\vspace*{4mm}

\begin{abstract}
We reanalyze the effective field theory (EFT) approach for the scenario in which the particles that account for the dark matter (DM) in the universe are vector states that interact only through the Standard Model-like Higgs boson. These DM particles are searched for in direct and indirect detection in astrophysical experiments and in invisible Higgs decays at the LHC. The constraints obtained in these two search types are complementary and correlated. In recent years, it has been advocated that the EFT approach is problematic for small DM mass and that it does not capture all the aspects of vector DM; one should thus rather interpret the searches in ultraviolet complete theories that are more realistic. In this note, we show that a more appropriate definition of the EFT with the introduction of an effective New Physics scale parameter, can encompass such issues. We illustrate this by matching the EFT to two examples of ultraviolet completions for it: the U(1) model with a dark photon and a model that was recently adopted by the LHC experiments in which vector-like fermions generate an effective interaction between the Higgs and the DM states at the one-loop level. Additionally, we find that the region of parameter space that is relevant for DM phenomenology is well inside the range of validity of the EFT. It thus provides a general parametrization of the effects of any ultraviolet model in the regime under exploration, making it the ideal framework for model-independent analyses of the vector DM Higgs-portal.
\end{abstract}

\newpage

\section{Introduction} 

The search for the weakly interacting massive neutral particles that form the Dark Matter (DM) in the Universe \cite{Bertone:2004pz,Arcadi:2017kky} is one of the main items in the program of various astroparticle and collider physics experiments. This is particularly true in the class of models in which these DM particles interact only through their couplings with the Higgs sector, the so-called Higgs-portal DM models \cite{Arcadi:2019lka} as they can be probed not only through Higgs exchange in DM detection in astrophysical experiments but also at the CERN LHC in invisible decays of Higgs bosons. In the simplest case where only the already observed Higgs particle \cite{Aad:2012tfa,Chatrchyan:2012xdj} is present, the Standard Model (SM) can be minimally extended by a DM particle that is isosinglet under the electroweak group but which can have the three possible assignments spin-0, spin-1 and spin-$\frac12$. In the scalar, vector  and fermionic cases, a rather simple effective field theory (EFT) approach can be adopted to describe the properties of the dark particle  \cite{Kanemura:2010sh,Djouadi:2012zc,Djouadi:2011aa} with the advantage of having only two extra parameters in addition to the SM ones, namely the mass of the DM state and its coupling to the Higgs boson. 

Such a minimal EFT has been investigated extensively and its predictions have been compared with the results of e.g. direct DM detection in experiments  such as  XENON1T \cite{Aprile:2017aty} and in searches for Higgs decays into DM pairs at the LHC \cite{ATLAS:2023tkt,CMS:2022qva}. In fact, it turns out that the LHC constraints on the DM states from the invisible Higgs decay width can be directly compared to the limits on the elastic  scattering cross section of the DM with nuclei obtained in  experiments such as XENON1T. The ATLAS and CMS collaborations are currently making extensive use of this correlation in the interpretation of their results. 

Alas, as any other EFT, this approach has a built-in shortcoming:  the Higgs-portal models are not complete in the ultraviolet (UV) regime in both the fermionic and vector DM cases. UV-completeness usually involves additional degrees of freedom which makes it so that, in principle, the complementarity between astroparticle and collider constraints cannot be established in the most general case. It was concluded that, at least in the vectorial DM case,  the simple and model-independent EFT approach cannot be used and, instead, one has to investigate the specific and more  complex UV-complete models. 

This problem might seem to be particularly severe in the vector DM case \cite{Baek:2012se,Baek:2014jga,Baek:2021hnl} for two reasons. First, for constant Higgs-DM coupling, the EFT becomes unreliable at small DM masses $M_V$. This is because perturbative unitarity is broken at an energy scale that depends linearly on $M_V$. Second, for such mass values, the invisible Higgs decay width becomes much larger than experimentally allowed.  Thus, even in the limit in which the new degrees of freedom of the UV-complete theory can be made phenomenologically irrelevant and the theory provides the same description as the EFT, which has been shown to be indeed the case in Refs.~\cite{Arcadi:2020jqf,Arcadi:2021mag}, it could appear that the EFT approach in the vector DM case should not be used in principle \cite{Baek:2012se,Baek:2014jga,Baek:2021hnl}. {This argument has been endorsed by the LHC collaborations which, in their latest analyses of invisible Higgs decays and their correlation with DM experiments, removed the case of the EFT with vector DM \cite{Aaboud:2019rtt,Khachatryan:2016whc,CMS:2022qva,CMS:2023sdw} in the comparison or included also some UV-complete cases  for illustration \cite{ATLAS:2023tkt}.}

In this note, we show that these problems do not arise on a closer examination of the EFT. The key point is that the DM-Higgs-portal coupling $\lambda_{HVV}$ is not independent of the new physics scale $\Lambda$. In fact, $\lambda_{HVV} \propto M_V^2/\Lambda^2$, making it clear that, if the new physics scale is kept constant and high enough  (e.g. at a few TeV), the coupling $\lambda_{HVV}$ must be  small when $M_V$ is small. In this way, the two problems at small DM mass mentioned above are avoided, since in the limit of small $M_V$ both the invisible Higgs width and the cross sections of the DM processes will be under control through $1/\Lambda^4$ dampening factors.

We illustrate this feature by matching the EFT to the two examples of ultraviolet complete models that are now used as benchmarks by the LHC collaborations: the U(1) gauge model \cite{Lebedev:2011iq,Baek:2012se,Falkowski:2015iwa} the DM being the dark photon and a model in which an effective Higgs-DM interaction is radiatively generated by vector-like fermions at the one-loop level \cite{DiFranzo:2015nli}.

\section{The EFT approach and its limitations}
\label{sec:old-eft}

The effective Higgs-portal scenario has been formulated for the spin-0, spin-$\frac12$ as well as for the spin-1 DM cases, see Ref.~\cite{Arcadi:2019lka} for a review. Assuming CP-conserving interactions, the latter case is described by the following effective Lagrangian \cite{Lebedev:2011iq,Djouadi:2012zc} 
\begin{eqnarray} 
\label{Lag:DM}  
{\cal L}_V = 
- \frac{1}{2} \partial_\mu V_\nu \partial^\mu V^\nu
+ \frac{1}{2} \partial_\mu V_\nu \partial^\nu V^\mu
+ \frac12  m_V^2 V_\mu V^\mu\! +\! \frac14
\lambda_{V}  (V_\mu V^\mu)^2\! +\! \frac14 \lambda_{HVV}  \Phi^\dagger \Phi
V_\mu V^\mu , 
\end{eqnarray} 
where $\Phi$ is the SM Higgs doublet field and $V_\mu$ the DM vector field; the $(V_\mu V^\mu)^2$ term is  not essential for our discussion and can be ignored. After electroweak symmetry-breaking, the original field $\Phi$ is decomposed as $(0, v + H)^T/ \sqrt{2}$ with $v=246$ GeV, inducing the trilinear interaction $HVV$ between the physical $H$ state  and DM pairs, and the DM mass term, 
\begin{eqnarray} 
M_V^2 = m_V^2 + \frac{1}{4}\lambda_{HVV} v^2 \, . 
\label{massV}
\end{eqnarray} 
The vector Higgs-portal model formulated above is extremely simple and predictive, featuring only $M_V$ and $\lambda_{HVV}$ as free parameters. For such a reason, together with its counterparts with spin-0 and $\frac12$ DM, is a popular benchmark for experimental studies. The $[M_V,\lambda_{HVV}]$ plane is mostly constrained by DM direct detection, with the world leading limit provided at the moment by the LZ collaboration \cite{LZ:2022lsv} (notice a very similar limit given as well by XENONnT \cite{XENON:2023cxc})  and, in the regime $M_V \leq \frac{1}{2}M_H$, by searches of invisible decays of the SM-like $H$ boson which set the 95\%CL constraint BR$(H \rightarrow VV) \equiv {\rm BR} (H\to {\rm inv}) \lesssim 10.7\%$ set by the ATLAS collaboration \cite{ATLAS:2023tkt}. An additional constraint can be achieved by requiring that the $V$ candidate accounts for the whole DM component of the Universe, generated via the freeze-out mechanism. This latter constrain should be taken with more grain-of-salt as it might be overcome by considering e.g. modified cosmological scenarios or assuming that the DM state is not stable on cosmological scales and/or does not account for the entire DM in the Universe; see Ref.~\cite{Arcadi:2019lka} for a discussion. Finally, the DM annihilation processes into SM fermions might occur at present times with a rate capable of producing experimentally detectable indirect signals. The corresponding limit are usually subdominant with respect to direct detection experiments and, hence, will not be reported explicitly in this work. 

Nevertheless, two arguments have been raised against this  scenario. We summarize them below referring to the original references~\cite{Baek:2012se,Baek:2014jga,Baek:2021hnl} for a more detailed discussion. 

\begin{figure}[!ht]
\vspace*{1mm}
    \centering
\subfloat{\includegraphics[width=0.47\linewidth]{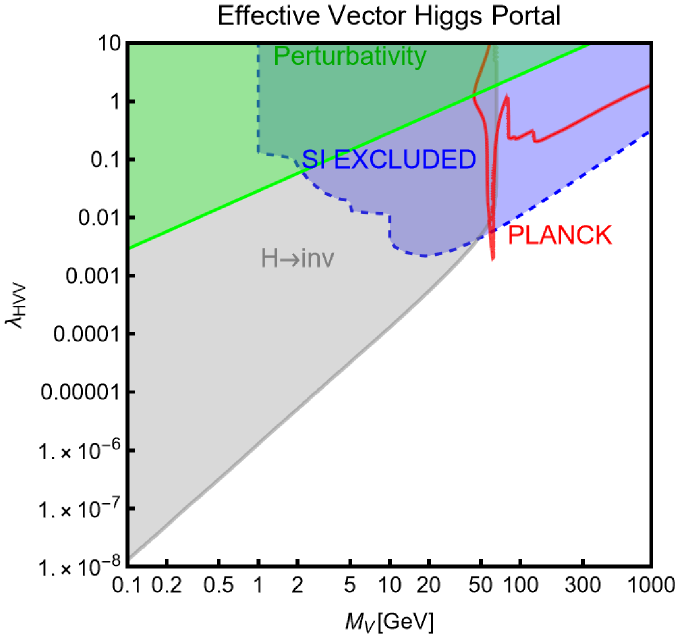}} ~~~   \subfloat{\includegraphics[width=0.48\linewidth]{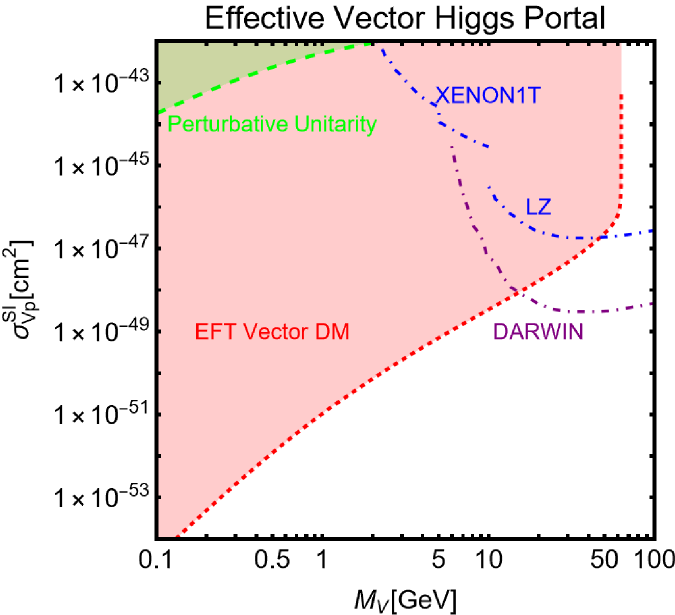}}
    \caption{Left panel: combination of DM constraints for the EFT  Higgs-portal for vector DM in the $[M_V,\lambda_{HV V}]$ plane. The gray region corresponds to BR($H \rightarrow \,\mbox{inv})>0.10$ \cite{ATLAS:2023tkt} the green region is excluded by perturbative unitarity. The red isocontour corresponds to the correct relic density assuming the freeze-out paradigm while the blue regions is excluded by direct detection constraints.  Right panel: the correlation plot between collider and astrophysical searches for the EFT vector DM Higgs-portal. The red area corresponds to the region excluded by LHC, while the dot-dashed line correspond to the limit from LZ and the one for a light DM from XENON1T and the purple dot-dashed line corresponds to the projected sensitivity of the DARWIN experiment. Again, the region marked in green represents the one excluded by the perturbative unitarity constraint.}
    \label{fig:pEFTV_updated}
\vspace*{-1mm}
\end{figure}

A first argument concerns  the invisible decay rate of the SM-like Higgs boson into DM pairs, $\Gamma_{\rm inv}= \Gamma(H \to VV)$, which can be written in the EFT approach as
\begin{align}
\Gamma_{\rm inv}|_{\rm  EFT}
=
\frac{\lambda_{HVV}^2 v^2 M_H^3}{512 \pi M_V^4}
\left(
    1
    - \frac{4 M_V^2}{M_H^2}
    + 12 \frac{M_V^4}{M_H^4}
\right)
{\left(
    1
    - \frac{4 M_V^2}{M_H^2}
\right)}^{1/2} \, . 
\label{Gamma-inv} 
\end{align}  
As it is clear from the previous equation, in the EFT approach where $\lambda_{HVV}$ and $M_V$ are in principle independent parameters, the Higgs partial decay width into DM can be very large and even diverges in the limit $M_V \rightarrow 0$. This results from the effect of the longitudinal degrees of freedom of the vector DM as will be discussed in some detail later. The LHC constraints on the invisible branching ratio then translate into a limit on the coupling $\lambda_{HVV}$ for each value of $M_V$. The excluded region in the $[M_V, \lambda_{HVV}]$ plane is displayed as the gray area in the left panel of Fig.~\ref{fig:pEFTV_updated}. The constraint is extremely strong\footnote{
It is noted in Ref.~\cite{Baek:2021hnl} that when the (unphysical) mass parameter $m_V$ in eq.~(2) vanishes, one would have the relation $\lambda_{HVV}=2M_V/v$ between the mass of the DM and its coupling to the Higgs boson. In this case, the invisible Higgs decay width becomes constant and in the limit $m_V=0$ will be given by $\Gamma_{\rm inv}^{\rm  EFT} = {M_H^3}/ {(32 \pi v^2)} \approx 320$ MeV, which is (by far) excluded by the LHC experiments. This simply means that while this  option is allowed by perturbative unitarity, it is excluded experimentally.} and excludes Higgs-DM couplings below the value $\lambda_{HVV} \approx 10^{-5}$ for DM masses below $M_V \approx 5$ GeV.  

A second argument raised in  Ref.~\cite{Baek:2021hnl} is that the vectorial DM Higgs-portal scenario is subject to the theoretical constraints of perturbative unitarity on scattering processes.  One of such processes\footnote{A similar problem occurs e.g. in associated production with a $Z$ boson of an off-shell Higgs boson which then  decays into two DM particles. Again, because of the longitudinal $V$ component, the cross section 
is inversely proportional to the mass $M_V$ \cite{Baglio:2015wcg,Ko:2016xwd} and, hence, would violate unitarity for very small values. } is $VV \rightarrow VV$ which proceeds via $H$-boson  exchange for which one can obtain the simple relation between the DM mass and $HVV$ coupling \cite{Lebedev:2011iq}
\begin{equation}
    M_V \geq \frac{\lambda_{HVV}v}{\sqrt{16\pi}} \label{eq:PUbound} \,  \ \ \  {\rm or} ~~
    \lambda_{HVV} \leq \sqrt{16 \pi} \frac{ M_V}{v} \, ,  
\end{equation}
which forbids low DM masses $M_V$ at large $\lambda_{HVV}$ couplings.  This is also exemplified in the left panel of Fig.~\ref{fig:pEFTV_updated} in which the green area shows the region that is excluded by the perturbative unitarity constraint. One can see, in particular,  that the effective Higgs-portal with a vector DM does not provide a consistent description for masses $M_V \lesssim 50 \,\mbox{GeV}$ when $\lambda_{HVV} \gtrsim 1$. This constraint is, however, relaxed for smaller Higgs couplings and masses $M_V \approx 1 \,\mbox{GeV}$ would be allowed for $\lambda_{HVV} \gtrsim 0.1$. For extremely small mass values very tiny $\lambda_{HVV}$ couplings would be required. Note that this theoretical constraint is superseded, by far, by the experimental one from the BR($H\to {\rm inv})$ measurement at the LHC. 

In the left panel of Fig. \ref{fig:pEFTV_updated}, together with the Higgs and theoretical constraints, we have also included some DM constraints. The red isocontour corresponds to the area in which the correct DM relic density, as measured by the Planck satellite \cite{Aghanim:2018eyx}, is achieved. The blue region corresponds to the exclusion from DM direct detection experiments, which has been obtained by combining the current world leading limits on spin-independent interaction given by the LZ experiment \cite{LZ:2022lsv} and by a dedicated analysis of the XENON1T experiment \cite{XENON:2019gfn}. 

In the EFT Higgs-portal, the invisible branching fraction of the $H$ boson and the DM scattering cross-section over nucleons, are proportional to the same coupling $\lambda_{HVV}$. It is then possible to re-express the DM scattering cross section $\sigma_{Vp}^{\rm SI}$ in terms of the invisible Higgs branching fraction BR$(H\to VV) \equiv  \Gamma(H\to VV)/ \Gamma_H^{\rm tot}$ with $\Gamma_H^{\rm tot}$ being the total width. One then would have the following correlation  function:
\begin{align}
\label{sigma-EFT}
    & \sigma_{Vp}^{\rm SI}|_{\rm EFT}=32  \mu_{Vp}^2 \frac{M_V^2}{M_H^3} \frac{{\rm BR} \left(H\rightarrow VV\right) \Gamma_H^{\rm tot}}{ \left(1-{4 M_V^2}/{M_H^2}+12 {M_V^4}/{M_H^4}\right){\left(1-{4 M_V^2}/{M_H^2}\right)}^{1/2} } \frac{1}{M_H^4}\frac{m_p^2}{v^2}|f_p|^2 \,  , 
\end{align}
where $\mu_{Vp} = {M_V m_p}/{(M_V+m_p)}$ is the DM-proton reduced mass. By fixing the value of BR$(H\rightarrow VV)$ to its experimental bound, it is possible to draw a line in the $[M_V,\sigma_{Vp}^{\rm SI}]$ plane and compare it with the bounds from dedicated astrophysical experiments.

Such a correlation plot in the $[M_V, \sigma_{Vp}^{\rm SI}]$ plane is shown in the right panel of Fig.~\ref{fig:pEFTV_updated}. The red area corresponds to the region excluded by the invisible Higgs branching ratio constraint  BR($H \rightarrow\,\mbox{inv}) \lesssim 0.10$ \cite{ATLAS:2023tkt}. The dot-dashed lines correspond to the leading limit from the LZ experiment \cite{LZ:2022lsv} and the one from a dedicated XENON1T measurement \cite{XENON:2019gfn} to probe light DM particles, while the purple dot-dashed line corresponds to the higher sensitivity projected by the DARWIN experiment \cite{Aalbers:2016jon}. Here again, we display the region (marked in green) excluded by the perturbative unitarity constraint. It is important to notice that in this correlation, the coupling $\lambda_{HVV}$ does not appear. 

In view of this analysis, it is clear that the EFT is predictive and well inside its range of validity for the relevant region of parameter space. However, a more careful inspection of the EFT validity and limitations is required in order to address the two concerns raised above, about the low-mass limit and unitarity violation.
This can be done by examining the high-energy behaviour it generates for various observables.  It should be noted that this information is partially hidden in the expression of the Lagrangian eq.~\eqref{Lag:DM}, which has the appearance of a renormalizable theory, as it only contains operators of dimension 4 or less, and no explicit cut-off scales are present. Nevertheless, perturbative unitarity is violated at high energies and so the theory is not truly renormalizable.

This is well-known to be due to the growth in energy of the longitudinal polarization vector $\epsilon_L$ and propagator $\Delta$ of a vector state with four-momentum  $p=(E,|\mathbf{p}|)$
\begin{equation}
    \epsilon_L(p) = 
    \left(\frac{|\mathbf{p}|}{M_V}, 0, 0, \frac{E}{M_V}\right),
    \qquad
    \Delta_{\mu\nu}(p)
    =
    \frac{i}{p^2 - M_V^2} \left(
        \eta_{\mu\nu} - \frac{p_\mu p_\nu}{M_V^2}
    \right) \, . 
\end{equation}

The effective nature of the theory can be made more explicit using  the Stueckelberg formulation, which describes the same physics, and we briefly review it here. In this formulation, an additional scalar field $\theta$ is introduced together with a U(1) gauge symmetry, ensuring that the number of degrees of freedom of the theory is preserved. The gauge transformations $\omega$ act non-linearly on $\theta$, as
\begin{equation}
    V_\mu \to V_\mu - \partial_\mu \omega \, ,
    \qquad
    \theta \to \theta + M_V \omega \, .
\end{equation}
The Stueckelberg Lagrangian is constructed by replacing $V_\mu$ in the original Lagrangian with the gauge-invariant quantity
\begin{equation}
    {\cal V}_\mu = V_\mu - \partial_\mu \theta / M_V,
\end{equation}
and is thus gauge invariant itself. The normalization of $M_V$ in front of the gauge transformation for $\theta$ (and thus the $1/M_V$ in the replacement for $V_\mu$) is such that the kinetic term for $\theta$ is canonically normalized. The theory now contains effective operators of dimension larger than 4, with coefficients proportional to inverse powers of $M_V$.
For example:
\begin{align}
    \mathcal{L}_{HVV}
    =
    \frac{\lambda_{HVV}}{4} \, \Phi^\dagger \Phi \, {\cal V}_\mu {\cal V}^\mu
    &=
    \frac{\lambda_{HVV}}{4} \, \Phi^\dagger \Phi \, V_\mu V^\mu
    \!+\! \frac{\lambda_{HVV}}{2 M_V} \, \Phi^\dagger \Phi \, V_\mu \partial^\mu \theta
    \!+\! \frac{\lambda_{HVV}}{4 M_V^2} \, \Phi^\dagger \Phi \, \partial_\mu \theta \partial^\mu \theta \, . 
    \label{eq:eft-M}
\end{align}

In the Lorentz gauge $\partial_\mu V^\mu = 0$, the longitudinal polarization is removed from $V_\mu$ and the vector propagator becomes $i \eta_{\mu\nu} / (p^2 - M_V^2)$. It is then clear that all inverse powers of $M_V$ in amplitudes must come only from the Wilson coefficients that appear in them. That is, their growth at high energies can be directly read from the Lagrangian. For example, the fastest-growing contribution to the scattering amplitude for $V_L V_L \to H H$ will be proportional to the coefficient of the last term of eq.~\eqref{eq:eft-M}:
\begin{equation}
    \mathcal{M}\, (V_LV_L \to HH) \sim \lambda_{HVV} \frac{s}{M_V^2},
\end{equation}
so that the scale of perturbative unitarity violation must be $\Lambda \sim M_V / \sqrt{\lambda_{HVV}}$. For a perturbative coupling $\lambda_{HVV} \lesssim O(1)$, we have $\Lambda \gtrsim M_V$, which ensures the consistency of the theory, as it must be valid at least up to energies around the masses of the particles it contains.

A ``natural" value of $\lambda_{HVV}$ of order unity leads to a cutoff $\Lambda$ that is close to $M_V$. However, the separation between these two scales can be made arbitrarily large by making $\lambda_{HVV}$ small. It is then convenient to trade the free parameter $\lambda_{HVV}$ for the cutoff $\Lambda$ by replacing $\lambda_{HVV}$ by  ${M_V^2}/{\Lambda^2}$. Alternatively, one may keep an adimensional $\overline{\lambda}_{HVV}$ free parameter and make the dependence on the mass $M_V$ explicit by introducing $\Lambda$, fixing it at a typical new physics scale (e.g. $\Lambda = 1$ TeV) and defining
\begin{equation}
    \lambda_{HVV} = \overline{\lambda}_{HVV} \frac{M_V^2}{\Lambda^2} \, .
\end{equation}

The observables of the theory are the same in both formulations eq.~\eqref{Lag:DM} and eq.~\eqref{eq:eft-M}, so we can make use of the expressions we have computed for them above, and perform this replacement in them. The invisible decay width of the Higgs $\Gamma_{\text{inv}}$ turns out to be proportional to $1/ \Lambda^4$, and thus small for sufficiently large $\Lambda$ values.

In summary, as for any other effective theory, the EFT for the vector DM Higgs-portal scenario described here is not renormalizable and violates perturbative unitarity at some new physics scale $\Lambda$.
It also requires some care in the low-mass regime. In particular, the Higgs-DM couplings should be constrained to be small at low DM masses.  Nevertheless, the current limit provided by the LHC collaboration in their invisible Higgs searches is so strong and constrains the Higgs-DM coupling to be so small that we are very far from the regime in which the  EFT is not valid. 

Furthermore, as will be highlighted in the next sections, the shape of the correlation plot will not depend on this coupling and will remain approximately the same even if the EFT is traded against a more realistic model. Hence, the Higgs-portal EFT with vector DM remains a valid benchmark for LHC searches, just the same way as it is the case for scalar and fermionic\footnote{We note that in the fermionic DM$\equiv \chi$ case \cite{Arcadi:2019lka}, similarly to what was discussed above, the Lagrangian describing the Higgs-$\, \chi\chi$ interaction,  ${\cal L}_\chi = - \frac12 M_\chi \bar \chi \chi - \frac14 {\lambda_{H\chi\chi}\over \Lambda} \Phi^\dagger \Phi \bar \chi \chi$, is not  renormalizable and the effective coupling $\lambda_{H\chi\chi}$ is damped by a new physics scale $\Lambda$ which is always assumed to be close to 1 TeV.} DM particles.

\section{The U(1) ultraviolet completion} 
\label{sec:uv-model}

The solution proposed by the authors of  Refs.~\cite{Baek:2012se,Baek:2014jga,Baek:2021hnl} was to move to an ultraviolet complete realization.
This has the clear advantage of completely avoiding the problems outlined above, with the downside that the model independence of the EFT approach is obviously lost.
The proposed model in Refs.~\cite{Baek:2012se,Baek:2014jga,Baek:2021hnl} is the simplest and most economical one: that in which the vector DM couplings are generated by a single additional scalar state. We briefly review below the essential features of this scenario. 

Apart from the SM fields and the DM vector $V_\mu$, the model only contains a complex scalar field $S$, which is charged under a dark U(1) gauge symmetry whose gauge boson is $V_\mu$. The terms in the Lagrangian involving $V_\mu$ and $S$ are given by
\begin{align}
    \mathcal{L}_{S,V}
    &=
    - \frac{1}{4} V_{\mu\nu} V^{\mu\nu}
    + D_\mu S^\dagger D^\mu S
    - V_{H,S},
    \\
    V_{H,S}
    &=
    - \frac{\mu_H^2}{2} |\Phi|^2
    +\frac{\lambda_H}{4} |\Phi|^4
    - \frac{\mu_S^2}{2} |S|^2
    + \frac{\lambda_S}{4} |S|^4
    + \frac{\lambda_{HS}}{4} |\Phi|^2 |S|^2,
\end{align}
where $V_{\mu\nu} = \partial_\mu V_\nu - \partial_\nu V_\mu$ and $D_\mu S = \partial_\mu S - i \tilde{g} V_\mu S$, with $\tilde{g}$ being the dark U(1) coupling constant. After electroweak  symmetry breaking, $\Phi$ and $S$ acquire vacuum expectation values, $v / \sqrt{2}$
and $\omega / \sqrt{2}$ respectively. This generates a mass term with coefficient $m_V = \tilde{g} \omega$ for $V_\mu$ (we will see later that,  additionally, the physical mass $M_V$ will receive a contribution from electroweak symmetry breaking which will be small in the scenario we consider here).

The scalar field $S$ can be parameterized as with real fields $\theta$ and $\rho$. The angular component $\theta$ is the would-be Goldstone boson for the spontaneously broken dark U(1), while the radial component $\rho$ acquires a mass $M_\rho^2 = \lambda_S \omega^2$ and mixes with the Higgs state $H$. The mass matrix for $H$ and $\rho$ is diagonalized by the eigenstates $H_1$ and $H_2$ with masses and mixing angle given by:
\begin{equation} 
M_{H_1,H_2}^2=\lambda_H v^2+\lambda_S \omega^2 \mp \frac{\lambda_S \omega^2-\lambda_H v^2}{\cos(2 \theta)}\, , \ \ \  \tan 2 \theta = \frac{\lambda_{HS}v \omega}{\lambda_S \omega^2-\lambda_H v^2} \, , 
\end{equation}  
The lowest eigenstate $H_1$ can be identified with the Higgs boson observed at the LHC. Adopting the $(M_{H_2},\sin\theta, \lambda_{HS})$ set of free parameters,  the quartic couplings $\lambda_H$ and $\lambda_S$ are then fixed; their expressions, together with other details, can be found in Ref.~\cite{Arcadi:2019lka}. 
 
The couplings of $H_1$ and $H_2$ to SM fermions and gauge bosons are given by the following Lagrangian (we ignore the one  $\mathcal{L}_{\rm S}^{\rm tril}$ giving the couplings of the scalars among themselves)
\begin{align}
\label{eq:Lag-SM-tril}
    & \mathcal{L}_{\rm S}^{\rm SM}=\frac{1}{v} (H_1 c_\theta + H_2 s_\theta)\left(2 M_W^2 W^{+}_{\mu}W^{-\,\mu}+M_Z^2 Z_\mu Z^\mu -m_f \bar f f\right) \, ,
\end{align}
while the complete Lagrangian describing DM phenomenology becomes
\begin{eqnarray}
   \mathcal{L}\!=\!\frac12 {\tilde{g}M_V} \left(H_2 c_\theta \!-\!H_1 s_\theta \right)V_\mu V^\mu \!+\! \frac18 {\tilde{g}^2}\left(H_1^2 s^2_\theta \!-\!2 H_1 H_2 s_\theta c_\theta \! +\! H_2^2 c^2_\theta\right)V_\mu V^\mu \!+\! 
\mathcal{L}_{\rm S}^{\rm SM}\!+\! \mathcal{L}_{\rm S}^{\rm tril}  \, .   
\end{eqnarray}
 One can trade the Higgs-portal coupling $\lambda_{HS}$ with the dark gauge coupling $\tilde{g}$ using  
 \begin{equation} 
 \lambda_{HS}=\tilde{g}\sin 2 \theta \frac{M_{H_2}^2 - M_{H_1}^2}{4 v M_V} \, .
 \end{equation}
The relevant set of input parameters can be chosen to be the physical one $\left( M_V, \tilde{g}, \sin\theta ,M_{H_2} \right)$. 
 
 Now, from perturbative unitarity in the processes $H_i H_i \rightarrow H_j H_j$, one obtains the constraint  $\lambda_i \leq \mathcal{O} \left({4 \pi}/{3}\right)$ \cite{Chen:2014ask} which when applied to $\lambda_{HS}$ constrains the hierarchy between $M_V$ and $M_{H_2}$: it might not be possible to have a very light DM and, at the same time, decouple $H_2$ from LHC phenomenology. This is one of the concerns raised by the LHC collaborations. It was addressed in Refs.~\cite{Arcadi:2020jqf,Arcadi:2021mag} where it was shown that one can have $M_{H_2} = {\cal O}(1\,{\rm TeV})$ together with a mixing angle $\theta$ that allows for measurable invisible $H_1$ decays.  

{Turning to the Higgs partial width into invisible DM pairs, its apparent divergence with the mass $M_V$ disappears since in the dark U(1) scenario, and contrary to the EFT, the mass of the DM and the coupling $\tilde{g}$ are not independent and one has $M_V=\tilde{g}\omega$. In this case, the limit $M_V\rightarrow 0$ is obtained my taking the limit $\tilde{g} \rightarrow 0$ when the new vev is non-zero $\omega \neq 0$\cite{Baek:2021hnl} (one could similarly also consider the limit $\omega \rightarrow 0$ for $\tilde{g}\neq 0$). In this case, the Higgs invisible partial decay width tends to a constant when $M_V=0$}
\begin{equation}
    \Gamma_{\rm inv}|_{\rm U(1)} \stackrel{M_V = 0 }{\longrightarrow} \frac{\tilde{g}^2 }{128 \pi}\frac{M_H^3}{\omega^2}\sin^2 \theta \, . 
\end{equation}
In this limit, $V$ can be simply seen as the Goldstone boson of the new U(1) symmetry.
 
In Ref.~\cite{Arcadi:2020jqf}, we have shown that the effective vector DM Higgs-portal can be obtained from this dark U(1) model by taking the limits $\sin\theta\! \ll\! 1$ and $M_{H_2} \! \gg \! M_{H_1}$ and that this limit is theoretically consistent with the unitarity constraint. Let us summarize the argument focusing  on the complementarity between DM direct detection  and invisible Higgs decay searches at the LHC, and not taking into account the issue of the cosmological relic density. 

Restricting to the case $M_{H_2}\!> \! M_{H_1}$ with $H_1\equiv H$, one obtains the following expressions for the invisible decay of the SM-like $H$ boson into $VV$ pairs, $\Gamma_{\rm inv}= \Gamma(H \to VV)$ and 
the vector DM spin-independent scattering cross section on protons $\sigma_{Vp}^{\rm SI}$ in the U(1) model
\begin{align}
\label{all-U1} 
&     \Gamma_{\rm inv}|_{\rm U(1)}=\frac{\tilde{g}^2 \sin^2 \theta}{128 \pi}\frac{M_{H_1}^3}{M_V^2} \left(1-{4 M_V^2}/{M_{H_1}^2}+12 {M_V^4}/{M_{H_1}^4}\right){\left(1-{4 M_V^2}/{M_{H_1}^2}\right)}^{1/2} \, , \\
& \sigma_{Vp}^{\rm SI}|_{\rm U(1)}=32  \mu_{Vp}^2 \frac{M_V^2}{M_{H_1}^3} \frac{{\rm BR}\left(H_1\rightarrow VV\right) \Gamma_{H_1}^{\rm tot}}{  \left(1-{4 M_V^2}/{M_{H_1}^2}+12 {M_V^4}/{M_{H_1}^4}\right){\left(1-{4 M_V^2}/{M_{H_1}^2}\right)}^{1/2}   } \frac{1}{M_{H_1}^4} \frac{m_p^2}{v^2}|f_p|^2 \kappa_{\rm U(1)} \, , \nonumber
\end{align}
with 
\begin{equation}
\label{eq:r_analytic}
\kappa_{\rm U(1)}= \cos^2 \theta \, M_{H_1}^4 \,  {\left(\frac{1}{M_{H_2}^2}-\frac{1}{M_{H_1}^2}\right)}^2 \, .  
\end{equation}
When $M_{H_2}\!>\!M_{H_1}\!=\!M_H$, one has $\Gamma_{H_1}^{\rm tot}=\Gamma_{H}^{\rm tot}$ for the total Higgs widths. Comparing eq.~(\ref{all-U1}) above with the corresponding ones in the EFT approach, eqs.~(\ref{Gamma-inv}) and (\ref{sigma-EFT}), one concludes that the predictions in the EFT and the dark U(1) model will coincide for a fixed invisible branching ratio BR$(H \rightarrow VV)={\rm BR}(H_1 \rightarrow VV)$ when in eq.~(\ref{eq:r_analytic}) one has 

In Ref.~\cite{Arcadi:2020jqf}, we have explicitly shown this by evaluating the ratio $r=\sigma_{Vp}^{\rm SI}|_{\rm U(1)}/\sigma_{Vp}^{\rm SI}|_{\rm EFT}$ in the plane $[M_V,\sigma_{Vp}^{\rm SI}]$ which is considered by the ATLAS and CMS collaborations in their analyses of the invisible Higgs decay branching ratio.  
We have performed a scan in the parameters $\sin\theta,M_{H_2},M_V$ in a wide range of their relevant values,  with the gauge coupling $\tilde{g}$ fixed in such a way as to obtain the desired  invisible branching ratio BR$(H_1 \rightarrow \mbox{inv})$.  We have then compared the obtained model points with the bounds from   direct detection experiments.  A comparison was also made with the predicted values in the EFT. 

The outcome of the analysis was that, indeed,  the EFT approach  can  represent a limiting case of the dark U(1) model,  while complying with perturbative unitarity constraints. 
As one might have expected, the EFT is a good approximation when $\kappa_{U(1)}$ is close to one, which requires, at the same time, that the mixing angle between the Higgs and the heavy scalar $\rho$ is small and that the mass $M_{H_2}$ of the extra physical scalar is large.

Hence, the simplest UV-complete model for a vector DM state with a Higgs-portal can be described by a simple EFT approach in which collider and astroparticle results can be consistently compared. In particular,  one is  able to  decouple in a consistent way the extra degrees of freedom in order to describe the model in the EFT approach for DM masses within the reach of colliders. In Ref.~\cite{Arcadi:2021mag}, it has been shown that this conclusion stays valid in more complex  UV-complete extensions in which either the gauge group is extended and/or  the symmetry-breaking mechanism of the hidden sector is made more complicated.

One can view this explicitly by matching the EFT to the UV model at the level of the Lagrangian. Acting on $S$, the non-linearly realized gauge symmetry we have described in section~\ref{sec:old-eft} for the EFT becomes the linear dark U(1) gauge symmetry. The relevant terms in the UV Lagrangian that generate the Higgs-portal in the EFT are 
\begin{equation}
    \mathcal{L}_{\text{UV}} \supset
    - \frac{1}{2} \rho (\square + M_\rho^2) \rho
    +    \rho \left(
        \frac{m_V^2}{\omega}  {\cal V}_\mu {\cal V}^\mu
        + \frac{\lambda_{HS} \omega}{4} \Phi^\dagger \Phi
    \right) \, .
    \label{eq:UV-lag-rho}
\end{equation}
The solution to the equation of motion for $\rho$, expanded in inverse powers of $M_\rho^2$ is then
\begin{equation}
    \rho =
    \frac{1}{M_\rho^2} \left(
        \frac{m_V^2}{\omega}  {\cal V}_\mu {\cal V}^\mu
        + \frac{\lambda_{HS} \omega}{4} \Phi^\dagger \Phi
    \right)
    + O\left(\frac{1}{M_\rho^4}\right) \, . 
\end{equation}
From this equation, it is clear that, when the electroweak symmetry is broken, $\rho$ acquires a vev, which generates a contribution to the physical mass $M_V$ of $V_\mu$ through eq.~\eqref{eq:UV-lag-rho}. This contribution is suppressed by $(v/M_\rho)^2$ with respect to $m_V$. Substituting this in the Lagrangian integrates out $\rho$, giving the following coefficient for the $HVV$ interaction:
\begin{equation}
    \mathcal{L}_{HVV}    
    \supset
    \frac{\overline{\lambda}_{HVV} m_V^2}{4\Lambda^2} \, \Phi^\dagger \Phi \, {\cal V}_\mu {\cal V}^\mu
    =
    \frac{\lambda_{HS} M_V^2}{4 M_\rho^2} \, \Phi^\dagger \Phi \, {\cal V}_\mu {\cal V}^\mu \ ,
    \label{eq:matching-correction}
\end{equation}
where we have used the parametrization $\lambda_{HVV} = \overline{\lambda}_{HVV} M^2_V / \Lambda^2$ introduced in section~\ref{sec:old-eft} that makes explicit the dependence on the DM mass $M_V$ and the new physics scale $\Lambda$ of the Higgs-portal coupling. From eq.~\eqref{eq:matching-correction} it is clear that the this dependence, which was derived from the EFT alone, matches the effects of integrating the extra degree of freedom in the UV completion. One may identify $\overline{\lambda}_{HVV} \simeq \lambda_{HS}$ and $\Lambda \simeq M_\rho$ in this case.

\begin{figure}[!ht]
\vspace*{1mm}
    \centerline{
\includegraphics[width=0.48\linewidth]{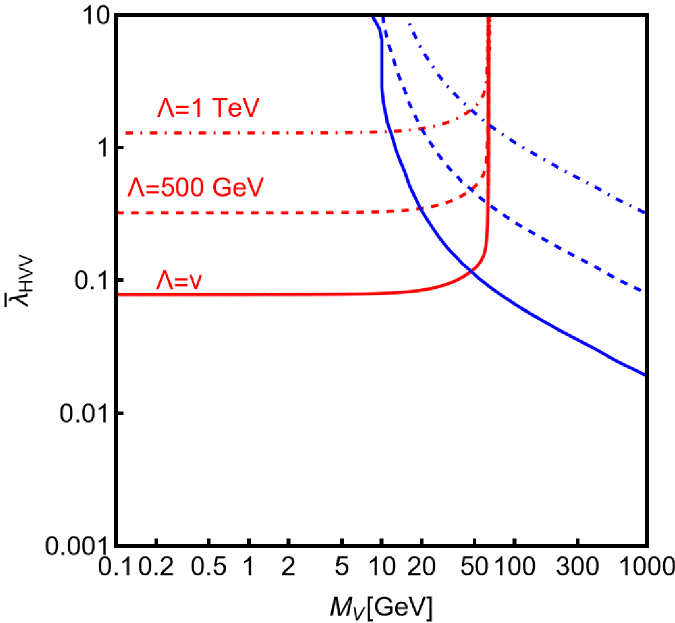}~
\includegraphics[width=0.48\linewidth]{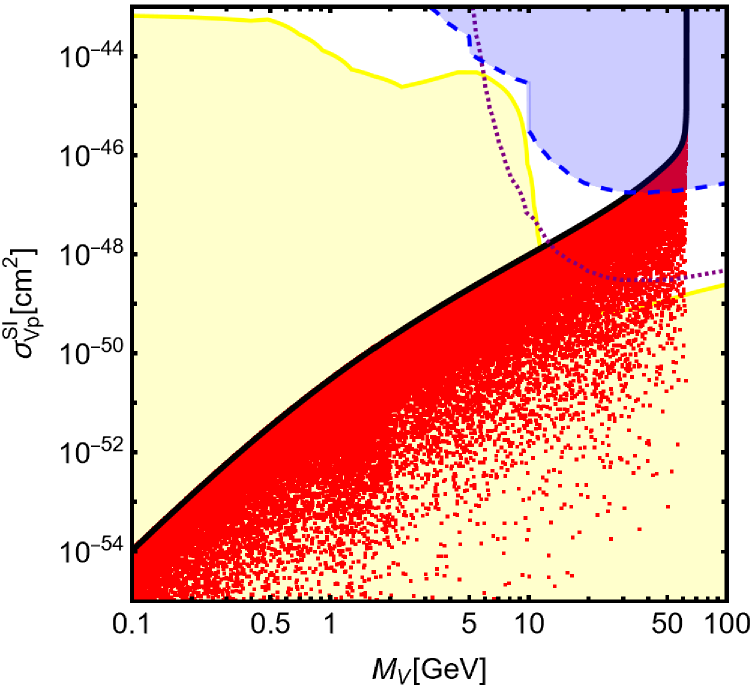}
 }
\caption{Left: constraints from the invisible Higgs branching ratio (red contours) and from DM direct detection constraints (blue contours) in the $[M_V, \bar \lambda_{HVV}]$ plane with three values of the scale ${\Lambda}$.  Right: values of the DM scattering cross section corresponding to BR$(H \rightarrow \mbox{inv})=0.10$, obtained by scanning over the dark U(1) model parameters space (red points).  These are compared with the prediction of the effective Higgs-portal (solid black line), the exclusion from LZ/XENON1T (blue region) and the expected sensitivities of DARWIN (dotted purple line). The yellow region corresponds to the neutrino floor.
}
    \label{Fig-scan}
\vspace*{-2mm}
\end{figure}

The use of the natural EFT parameter $\overline{\lambda}_{HVV}$ with the dependence on $M_V^2 / \Lambda^2$ stripped out leads to modifications in the exclusion contours in the $[M_V, \overline{\lambda}_{HVV}]$ plane when compared to those in $[M_V, \lambda_{HVV}]$.
This is shown on the left-hand side of Fig. \ref{Fig-scan} for three values of the scale $\Lambda$. In particular, the Higgs decay width does not feature any longer the dangerous scaling with the DM mass. As it can be expected, increasing the value of $\Lambda$ weakens the constraints on the coupling $\overline{\lambda}_{HVV}$. Adopting the rule of thumb $\overline{\lambda}_{HVV} \lesssim 1$, we see that the framework under consideration provides a viable phenomenology for $\Lambda \lesssim 1\,\mbox{TeV}$.

On the other hand, as already pointed, the correlation plot does not depend on the DM coupling with the Higgs boson $\lambda_{HVV}$ or $\overline{\lambda}_{HVV}$. The overall rescaling induced by our re-parametrization affects in equal footing the Higgs decay width and the DM scattering rate so that the correlation line does not depend on $\overline{\lambda}_{HVV}$ and maintains the exact shape as in the conventional EFT formulation of the vector Higgs-portal.

In order to provide a visualization of the validity of the EFT approximation, we have followed Ref.~\cite{Arcadi:2020jqf}, and evaluated the ratio $r=\sigma_{Vp}^{\rm SI}|_{\rm U(1)}/\sigma_{Vp}^{\rm SI}|_{\rm EFT}$ in the plane $[M_V,\sigma_{Vp}^{\rm SI}]$ which is considered in the LHC analyses using a scan in the parameter ranges $\sin\theta \in \left[10^{-3},0.3\right]$, $M_V \in \left[10^{-2},62.5\right]\,\mbox{GeV}$, $M_{H_2} \in \left[125.1,1000\right]\,\mbox{GeV}$,  with the gauge coupling $\tilde{g}$ fixed in such a way as to obtain a fixed invisible branching ratio BR$(H_1 \rightarrow \mbox{inv})$. We update here this analysis by assuming an invisible branching ratio of BR$(H_1 \rightarrow \mbox{inv})=10\%$ which is the present sensitivity \cite{ATLAS:2023tkt} and using the latest XENON1T results. The outcome is shown on the right-hand side of Fig.~\ref{Fig-scan}, where the obtained model points (in red) are compared with the exclusion bounds from XENON1T (blue region) and the projected sensitivities from the XENONnT \cite{Aprile:2018dbl} and DARWIN experiments \cite{Aalbers:2016jon}. A comparison is also made with the outcome of the invisible Higgs branching ratios in the EFT (thick solid black lines). As can be seen, while indeed the DM scattering cross section correlated to a fixed value of the invisible Higgs branching fraction can be orders of magnitude below the one expected in the EFT approach,  the latter  can represent a limiting case of the dark U(1) model. This is achieved consistently with the unitarity and perturbativity constraints discussed previously. 
 
\section{The radiative Higgs-portal model}
\label{sec:radiative}

Before closing, let us consider for completeness a radiative Higgs-portal model with vector DM states \cite{DiFranzo:2015nli} which has been included recently in the LHC analyses \cite{Zaazoua:2021xls,ATLAS:2023tkt,CMS:2023sdw} correlating the Higgs searches with direct detection experiments. In this model, the vector DM is, again, interpreted as the gauge boson of an extra U(1) symmetry. This time, however, it is assumed that the tree-level coupling between the DM and the SM Higgs is negligible. The contact between the hidden and the visible sector will be established by vector-like fermions charged under both the SM SU(2)$\times$U(1) and the new U(1) symmetries. In order to avoid the appearance of chiral anomalies, the minimal field content is represented by 6 two-component Weyl spinors with the following charges under $\rm SU(2) \times U(1)_Y \times U(1)^\prime$:
\begin{equation}
    \psi_{1,2} \sim (2,1/2,\pm 1),\qquad \qquad
    \chi_{1,2}\sim (2,-1/2,\mp 1), \qquad \qquad
    n_{1,2}\sim (1,0,\mp 1)\, . 
\end{equation}
The interaction Lagrangian with the Higgs field $H$ can be written as

\begin{align}
     \mathcal{L} \subset & -m \epsilon^{ab}\left(\psi_{1a}\chi_{1b}+\psi_{2a}\chi_{2b}\right)-m_n n_1 n_2\nonumber\\
    & -y_\psi \epsilon^{ab}\left(\psi_{1a}\Phi_b n_1+\psi_{2a}\Phi_b n_2\right)-y_\chi \epsilon^{ab} \left(\chi_{1a} \Phi^{*}_b n_2 + \chi_{2a} \Phi^{*}_b n_1\right)+\mbox{h.c.} \, . 
\end{align}
After symmetry breaking and the consequent mixing between the different fields, a physical spectrum made by three electrically neutral $N_i$ and two charged $E_{1,2}$ Dirac fermions arise. Their gauge interactions are described by the following Lagrangian ($e$ is the electric charge and $g=e/s_W$ the $SU(2)$ gauge coupling with $\theta_W$ being the Weinberg angle): 
\begin{align}
    & \mathcal{L}_{\rm gauge}=e \left(\bar E_1 \gamma^\mu E_1-\bar E_2 \gamma^\mu E_2\right) \left(A_\mu +\frac{(1-2 s_W^2)}{2 c_W s_W}Z_\mu\right)+\frac{e}{2 c_W s_W} \bar N_i \gamma^\mu G^Z_{ij}N_j Z_\mu \nonumber\\
    & +\frac{e}{\sqrt{2}s_W}\left[\left(\bar E_1 \gamma^\mu G_{i1}^{W} N_1+\bar N_i \gamma^\mu G_{i2}^W E_2\right)W_\mu^+ +\mbox{h.c.}\right]
     +g \left(\bar E_i \gamma^\mu E_i+\bar N_i \gamma^\mu N_i\right) V_\mu \, , 
\end{align}
with $V_\mu$ the vector DM field and $G_{ij}^{W,Z}$ parameters depending on the mixing among the different new fermionic states. The leading effects of the vector-like fermion loops on DM phenomenology can be described by the following effective interactions:
\begin{equation}
\label{eq:radlag}
    \mathcal{L}_{\rm loop}=- \frac{1}{4}A(p^2) H V^{\mu \nu}V_{\mu \nu} - \frac{1}{2}B(p^2)H V^\mu V_\mu\, . 
\end{equation}
with form factors $A(p^2)$ and $B(p^2)$ that have been computed analytically in Ref.~\cite{DiFranzo:2015nli}. The results (including the full dependence on the masses and momenta) are based on computation of these loop functions via the packages FeynCalc \cite{Mertig:1990an,Shtabovenko:2020gxv} and Package-X \cite{Patel:2015tea}.

It is instructive to consider the case in which the mass of the fermions is large compared to $M_V$ and $p$. Then, the interactions in eq.~\eqref{eq:radlag} become local, and can thus be incorporated into the local EFT described in section~\ref{sec:old-eft}. For simplicity, we assume the single-fermion limit described in Ref.~\cite{DiFranzo:2015nli}, in which the contribution from only one of the new fermions, 
with mass $m$ and Yukawa coupling $\overline{y}$, 
dominates the $A$ and $B$ functions. In this regime,
\begin{equation}
    A(p^2, M_V^2 \ll m^2)\approx -\frac{g^2 \overline{y}^2 v}{6 m^2},\,\,\,\,B(p^2, M_V^2 \ll m^2) \approx -\frac{g^2 \overline{y}^2 v M_V^4}{360 m^4}\, . 
    \label{eq:AB-approx}
\end{equation}
In order to keep the dependence on the relevant scales explicit, we have parametrized the Yukawa coupling $y$ employed in Ref.~\cite{DiFranzo:2015nli} as
\begin{equation}
    y = \frac{\sqrt{2} \overline{y}^2 v}{m}.
\end{equation}
The correctness of this scaling can be verified by examining the Feynman diagrams that contribute to the interactions under consideration in the unbroken phase.
To accommodate both types of interactions in the EFT, one needs to include in its Lagrangian the dimension-6 gauge-invariant operator $|\Phi|^2 V^{\mu\nu} V_{\mu\nu}$, in addition to the $|\Phi|^2 V^\mu V_\mu$ operator we have considered so far. 
The dimension of $|\Phi|^2 V^{\mu\nu} V_{\mu\nu}$ dictates that it has a coefficient of order $1/\Lambda^2$. 
No additional factors of $M_V^2 / \Lambda^2$ appear naturally in this case, in contrast with the situation for the $|\Phi|^2 V^\mu V_\mu$ operator, because $V^{\mu\nu} V_{\mu\nu}$ does not contain the problematic longitudinal polarizations.
Thus, we parametrize the coefficients of the interactions in the EFT as:
\begin{equation}
    \mathcal{L}_{HVV}    \supset
    \frac{\gamma}{\Lambda^2} |\Phi|^2 V^{\mu\nu} V_{\mu\nu}
    + \frac{\overline{\lambda}_{HVV} M_V^2}{\Lambda^2}
    |\Phi|^2 V^{\mu} V_{\mu} \, . 
\end{equation}
Using eqs. \eqref{eq:radlag} and \eqref{eq:AB-approx}, and identifying the new physics scale with the fermion mass $\Lambda = m$, we then obtain the matching conditions
\begin{equation}
    \gamma = -\frac{g^2 \overline{y}^2}{48}, \qquad
    \overline{\lambda}_{HVV} = -\frac{g^2 \overline{y} M_V^2}{1440 \Lambda^2}\, . 
\end{equation}
Therefore, the coefficient of the first operator has the scaling with $\Lambda$ derived within the EFT, while the second one is suppressed by an additional factor of $M_V^2 / \Lambda^2$.
At any rate, the EFT is able to capture the effects of the radiative model when $m$ is large compared to $p$ and $M_V$.

Regarding the observables of this radiative model, omitting the phase-space factors $\propto M_V^2/M_H^2$ for simplicity, the invisible Higgs width decay can be simply written as\footnote{Note that, for practical purposes, this expression for the invisible Higgs width in the radiative model is identical to the one generated by the $|\Phi|^2 V_\mu V^\mu$ portal interaction, given in eq.~(\ref{Gamma-inv}), with the identification for the coupling  $\overline{\lambda}_{HVV} \rightarrow {g^2 \overline{y}^2}/({6\pi^2})$ and the new physics scale $\Lambda = m$.}
\begin{align}
\Gamma_{\rm inv}|_{\rm rad} & \stackrel{M_H \ll m }{\longrightarrow}  \frac{ g^4 y^2 M_H^3}{4608\pi^5 m^2} \, .
     \label{eq:widthrad}
\end{align}
The correlation plot is more complicated to obtain as the DM scattering cross section on nucleons via $H$-exchange has a slightly different dependence on the loop induced couplings 
\begin{align}
   & \sigma_{Vp}^{\rm SI}|_{\rm rad}=\frac{\mu_{Vp}^2}{4 \pi}\frac{m_p^2}{M_H^4 M_V^2}\left(\frac{f_p}{v}\right)^2 \bigg\vert B(p^2\simeq 0)-M_V^2 A(p^2\simeq 0)\bigg\vert^2
    \approx \frac{\mu_{Vp}^2 g^4 y^2}{288\pi^5 }\frac{m_p^2 M_V^2}{M_H^4 m^2}\left(\frac{f_p}{v}\right)^2 \, . 
\label{eq:sigma-rad}
\end{align}
We stress again that the correlation between  $\sigma_{Vp}^{\rm SI}$ and  $\Gamma_{\rm inv}$ does not depend on the Higgs coupling to the DM particles and is exactly the same as in the EFT approach. In addition, by comparing eqs.~(\ref{eq:widthrad}) and (\ref{eq:sigma-rad}), one can see  that the correlation plot will not depend on the value of the mass scale $m$.

\begin{figure}
    \centering
    \subfloat{\includegraphics[width=0.45\linewidth]{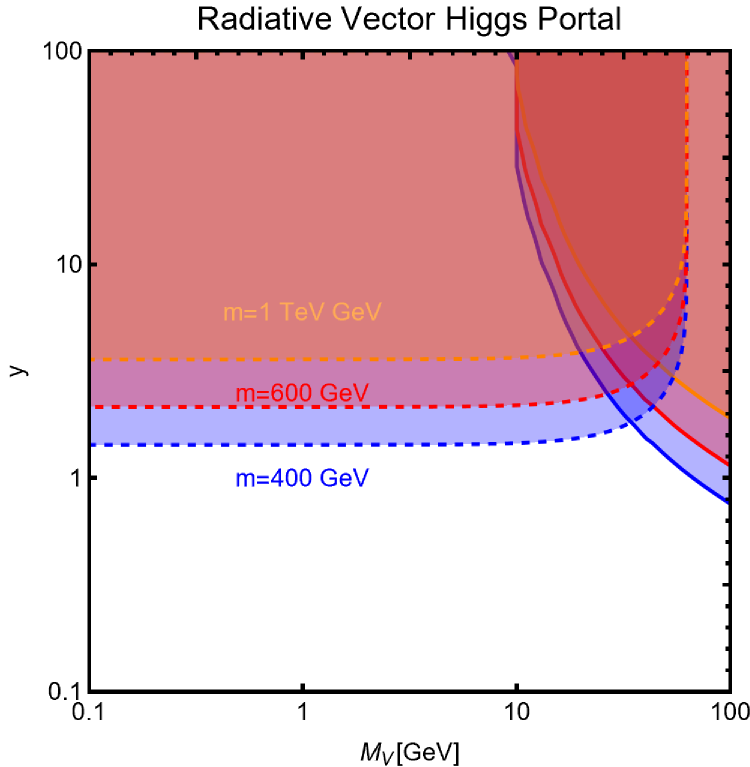}}~~
    \subfloat{\includegraphics[width=0.475\linewidth]{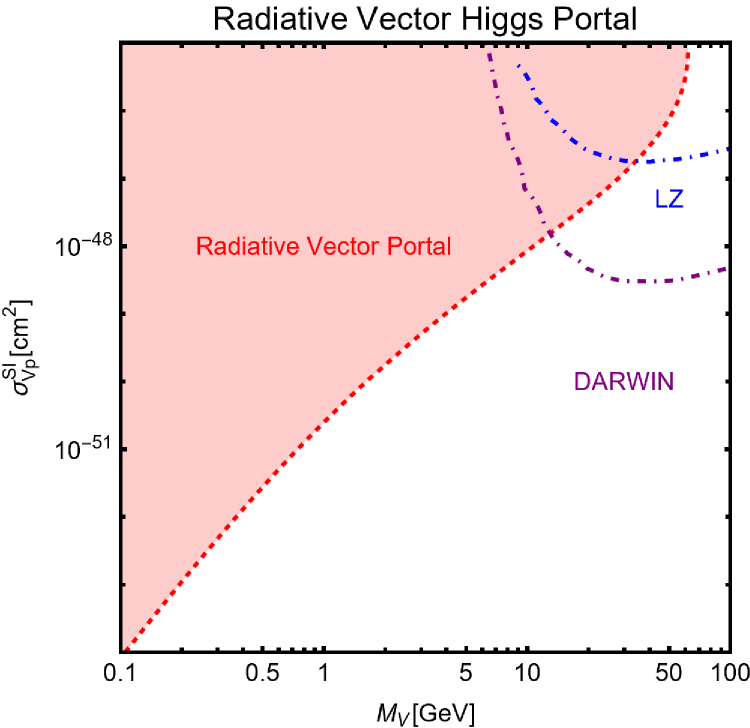}}
    \caption{{Left Panel}: Constraints in the $[M_V, y]$ plane from the Higgs invisible width and from DM direct detection experiments (the regions above the lines are excluded). The different colors corresponds to different mass scales, namely 400 GeV (blue), 600 GeV (red) and 1 TeV (orange), of the heavy fermion which is integrated out to obtain the effective DM-Higgs couplings.  Right Panel: the invisible Higgs decay branching ratio at the LHC and direct DM detection correlation plot in the radiative model; the constraints from the DARWIN and LZ experiments are also included.}
    \label{fig:radportal}
\end{figure}

A summary of the results of the radiative model is given in Fig.~\ref{fig:radportal}. The left panel shows the constraints from the Higgs invisible width and direct DM detection in the $[M_V,y]$ plane, again assuming for simplicity, a  single fermion with a Yukawa coupling $y$ running in the loop\footnote{We remark that in the approximation of a dominance of a single new fermion in the loop, one would need a very large and  possibly non perturbative Yukawa coupling to obtain a Higgs decay width and a DM scattering cross section close to the experimental sensitivity. A more realistic realization would then involve the presence of several, possibly close in mass, new fermions.}. The different colors correspond to three assignments of the new fermion mass, namely $m=400$ GeV, 600 GeV and 1 TeV. The right panel shows, instead, the customary correlation plot between collider and astrophysical constraints. Here also, while the expression eq.~(\ref{eq:sigma-rad}) for the DM scattering cross section differs from those obtained in the other models discussed in this work, the $M_V \rightarrow 0$ limit strongly resembles the one of the EFT defined in section 3 with an additional $|\Phi|^2 V^{\mu\nu} V_{\mu\nu}$ operator. 

\section{Conclusions}

We have addressed two concerns and potential problems which have been raised about the use of the effective field theory description of vector DM particles that interact with the Standard Model through the Higgs portal.
These concerns prevented the ATLAS and CMS collaborations to use the EFT as a benchmark to correlate their searches for invisible decays of the SM-like Higgs boson at the LHC with constraints set by direct DM detection in astrophysical experiments. The two problems are related with the limit of small DM mass, $M_V \to 0$ which, if care is not taken, would lead to a violation of perturbative unitarity and to an invisible Higgs decay width that is far too large to be compatible with experiment. 

We have shown that, on a closer inspection of the EFT, the two problems are not present, if the dependence of this coupling on the new physics scale $\Lambda$ is taken into account. We have found that $\lambda_{HVV} \propto M_V^2/ \Lambda^2$ so that, even for a very light DM state, the coupling stays small when the new scale $\Lambda$ is of the order of the TeV and the DM mass $M_V^2$ is well below it. In this case, both the invisible Higgs decay branching ratio and the DM cross sections, including those that could lead to unitarity violation, are kept under control. 

We have exemplified this feature by matching the EFT to two ultraviolet complete scenarios with vector DM coupled to the Higgs boson. Both of them have been used by the LHC collaborations, instead of the EFT approach,  to correlate their results with those from direct detection experiments. A first one (also discussed in a previous analysis \cite{Arcadi:2020jqf}, that we update here)  is the  celebrated case in which the vector DM couplings are generated by a single additional scalar state which is charged under a U(1) dark gauge symmetry whose gauge boson is the vector DM. Another scenario, recently adopted by the ATLAS and CMS collaborations,  is a radiative Higgs-portal model in which the vector DM is also interpreted as the gauge boson of an extra U(1) symmetry but its coupling to the Higgs boson is generated only at the one-loop level by vector-like fermions that are charged under this symmetry. In both cases, we have shown that the EFT can represent a very good limiting case of these UV-complete models while being consistent  with unitarity constraints and those on the invisible Higgs branching ratio. This is particularly true for the correlation between the results of the LHC and astrophysical experiments, since it does not depend on  the Higgs-DM coupling.

This result is not really surprising.
Indeed, the EFT we have considered, as any other EFT (in the DM context, see e.g. the discussion in Ref.~\cite{Criado:2021trs}), is able to capture the effects of any of its UV completions at sufficiently low energies.
One thus gets a UV-model-independent description of the physics under consideration, at the price of having pertubative unitarity violation at high energies.
In practice, we have found that the relevant region of parameter space for DM phenomenology is always in the regime in which the effective approach is valid.\bigskip

\noindent {\bf Acknowledgements:}
The authors warmly thank Robert Ziegler for early collaboration and fruitful discussions. G.A. thanks the University of Granada for the warm hospitality during part of the completion of this work. AD is supported by the Junta de Andalucia through the Talentia Senior program and the grant PID2021-128396NB-I00. J.C.C. is supported by grant RYC2021-030842-I funded by MCIN/AEI/ 10.13039/501100011033 and by NextGenerationEU/PRTR, and grant PID2022-139466NB-C22 funded by MCIN/AEI and by ERDF.

{\small 

\setstretch{.9}
\bibliographystyle{unsrt}
\bibliography{vdm-biblio}
}

\end{document}